# A New Approach to Enhance Security of Visual Cryptography Using Steganography (VisUS)

Uttam Kr. Mondal[#], Shamayita Pal[&1], Amit Ranjan Dutta[&2], J.K.Mandal[*]

[#]*Dept. of CSE & IT,College of Engg. & Management, Kolaghat , Midnapur(W.B), India.*
`uttam_ku_82@yahoo.co.in`
[&]*4th year, Dept. of CSE, College of Engg. & Management, Kolaghat , Midnapur(W.B), India*
[1]`shamayitacse.cemk@gmail.com`
[2]`amitcemk.cse@gmail.com`
*[*]Dept. of CSE, University of Kalyani, Nadia(W.B),India*
`jkm.cse@gmail.com`

*Abstract*— **Steganography is a process that hides secrete message or secrete hologram or secrete video or secrete image whose mere presence within the source data should be undetectable and use for transmitting secret information over public media. Visual cryptography is a cryptographic technique in which no cryptographic computation is needed at the decryption end and the decryption is performed by the human visual system (HVS). In this paper, both Steganography and visual cryptography have been selected to provide more secure data transmission over the public media with less hazard of computation. This technique generates shares with less space overhead as well as without increasing the computational complexity compared to existing techniques and may provide better security. It is also easy to implement like other techniques of visual cryptography. Finally, experimental results are given to establish the security criteria.**

## I. INTRODUCTION

Visual cryptography, introduced by Naor and Shamir in 1995 [2], is a new cryptographic scheme where the ciphertext is decoded by the human visual system (HVS) without any complex cryptographic computation . Using the visual cryptographic computation , any text or image to be encrypted is fed as an image (as the input) in the system to generate 'n' number of (where n is a positive integer grearer than or equal to 2) different output images (called shares). A share looks like an image of some random noises. For decryption the recipient has to stack a minimum number of shares(printed in transparencies) in an arbitrary number with the proper alignment . In this paper we have used (n,n) visual cryptography where the system generates n (n ≥ 2) number of shares and at least any k (2 ≤ k ≤ n) shares are needed to regenerate the secret information. On the otherside, steganographic [3, 4] is used to hide data may be secrete message or secrete hologram or secrete video or secrete image whose mere presence within the source data should be undetectable. Hiding a message/image inside an image without changing its visible properties [5] and by altering source image is a challenging task. The least-significant bit (LSB) replacement developed by Chandramouli et al. [3] by masking, filtering and transformations on the source image is a common method to make these alterations. Dumitrescu et al. [6] proposed an algorithm for detecting LSB Steganography.

In this present paper, a new technique has been proposed which is used to hide some secret information in some specific positions of a cover image without changing its visual appearance. Then, the cover image will transfer over public media by creating a number of shares. Even, all shares are not needed to reconstruct the cover image at receiving end.

Section 2 of the paper deals with the background theory of cryptography technique. The proposed technique has been depicted in section 3. Experimental results are given in section 4. Securities level testing for the proposed algorithm is made in section 5. Conclusions are drawn in section 6. References are given at end.

## II. THEORY

### A. Image Based Steganography

The steganography techniques require two files: cover media that helps to hide the data, and the data itself.

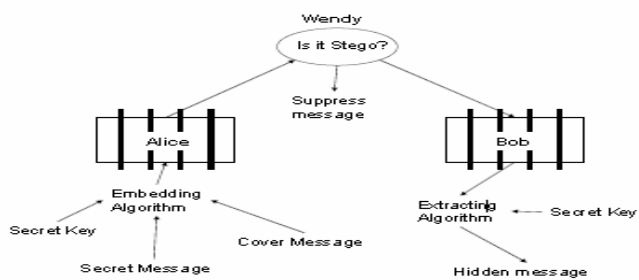

Fig 1 General Model of Steganography

Here the cover media is an Image and the Data is a text message. After embedding the message in the cover image file, the newly produced image is called Stego image. As this technique is mainly used in public media, We use an embedding algorithm or encoding algorithm at sender site and the extracting or decoding algorithm at receiver site[1].

*B. Visual Cryptography*

Visual Cryptography deals with secret sharing of data. Here an image is splitted into **shares**. The basic model for visual sharing of the *k* out of *n* secret image is such that;

- Any *n* participants can compute the original message if any *k* (or more) of them are stacked together.
- No group of *k-1* (or fewer) participants cannot compute the original message.

Images are split into two or more shares such that when a predetermined number of shares are aligned and stacked together, then the secret image is revealed [2] without any computation.

*1) (n, n) Visual Cryptography*

In this type of visual cryptographic scheme, the system generates n (n ≥ 2) number of shares and all shares are needed to be stacked together to get back the secret information.

*2) (k, n) Visual Cryptography*

In this type of visual cryptographic scheme, the system generates n (n ≥ 2) number of shares and at least any k (2 ≤ k ≤ n) shares are needed to regenerate the secret information.

Various algorithms [7, 8, 9, 10] are available for different visual cryptographic schemes, where efforts have been made to enhance the security. From the literature it can also be traced that efforts has also been made to increase the ease of use of the visual cryptography.

### III. THE TECHNIQUE(VisUS)

*A. Image Steganography*

*1) Text Message Encryption*

**Input**: A cover image, text message.

**Output**: A stego image.

The embedding technique of text into a cover image are given in the following step 1 to 5.

- Step 1: Insert message only on those pixels of cover image where the R, G, B values are less than 40 and make them equal to 40.
- Step 2: Represent the text message into binary stream and make groups of two bits. If "00011100" is binary stream then 00,01,11,00 respectively are groups of two bits.
- Step 3: Represent each group using equivalent decimal number like 00 will represented by 0, 01 by 1, 10 by 2, 11 by 3 respectively.
- Step 4: Add the equivalent decimal number with any of R,G,B value to the pixels getting from step 1 sequentially.
- Step 5: Continue step 4 until all groups are added with cover image.

*B. Applying Visual Cryptography*

*1) Sharing*

This approach works in bit level visual cryptography, i.e. instead of block wise sharing we used **BIT WISE SHARING.** One image is divided into **7 shares** where in every share at every pixel position some bit values are missing**.** so it is impossible to get the complete information from one share. Until **any 4 of those shares** are stacked properly nobody will get the proper image. The method of sharing is graphically presented in figure 2 and the bit patterns of shares are given in table 1.

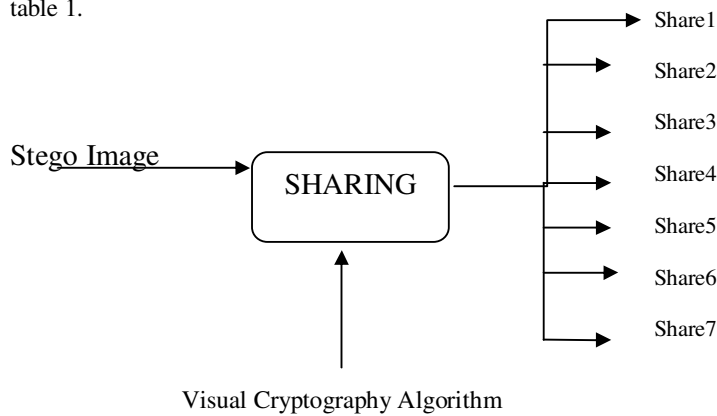

Fig 2 Sharing process in visual cryptography

TABLE 1

IMAGE SHARES

| share | 8th bit | 7th Bit | 6th bit | 5th bit | 4th bit | 3rd bit | 2nd bit | 1st bit |
|---|---|---|---|---|---|---|---|---|
| Share1 | √ | √ |   | √ |   | √ |   | √ |
| Share2 | √ |   | √ |   | √ |   | √ | √ |
| Share3 | √ | √ |   |   | √ | √ | √ | √ |
| Share4 | √ |   | √ | √ |   |   | √ |   |
| Share5 | √ | √ |   | √ | √ |   |   | √ |
| Share6 | √ |   | √ | √ |   | √ | √ |   |
| Share7 | √ | √ | √ |   | √ | √ |   |   |

√→shows present bits

*C. Reconstruction*

*1) Reconstructing the stego image*

Reconstructing the original stego image from any 4 share images is very simple. No computation process is needed, only BITWISE OR operation will be performed and the whole image will be reconstructed without any loss.

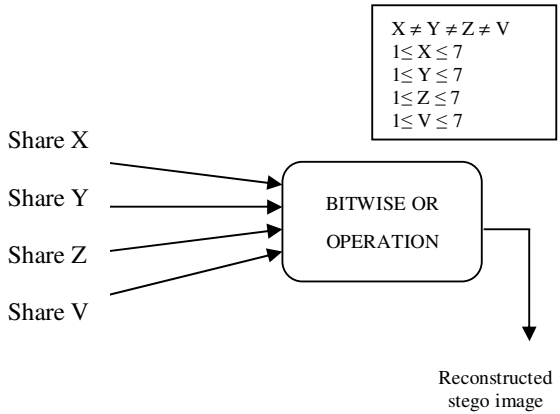

Fig 3 Reconstructing image using shares

*2) Separating text message*

Retrieving text message from stego image are given in the following step 1 to 3.

Step 1: search all pixels of stego image where any two of the R, G, B values are equal to 40 and other value is equal to any one of the set {40,41,42,43}.

Step 2: Find the extra value from those pixels and represent them into equivalent binary patterns. Where 0 will be represented by 00 , 1 by 01,2 by 10 and  3 by 11 respectively.

Step 3: Continue step 2 for all pixels which are found after applying step 1 and put them sequentially for getting the binary pattern of text message.

IV. EXAMPLE

Taking a standard image i.e. Image-1 in bitmap format (figure 4) as a cover image and after inserting a text message which is shown in figure 4 to 5.

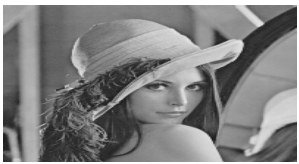

Fig 4: Image-1(Cover Image)

Text message : "Steganography"

After inserting above this text message ,the generated image is Image-2, shown in figure 5.

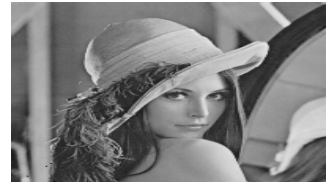

Fig 5: Image-2(Stego Image)

Generated shares after applying above method of visual cryptography are shown in figures 6 to 12.

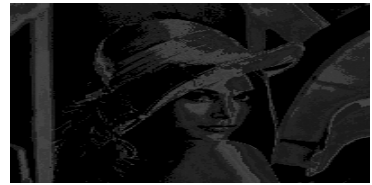

Fig 6:  Share 1

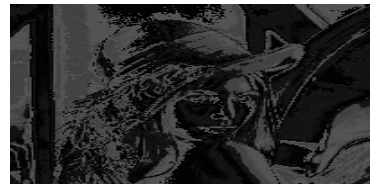

Fig 7:  Share 2

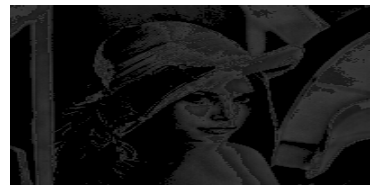

Fig 8:  Share 3

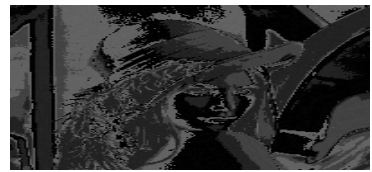

Fig 9:  Share 4

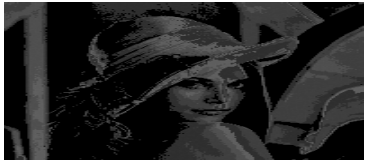

Fig 10: Share 5

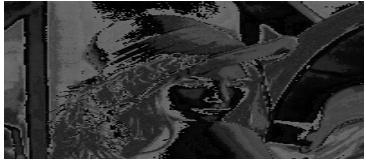

Fig 11: Share 6

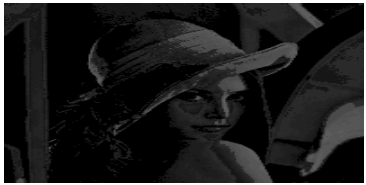

Fig 12: Share 7

Any 4 of above shares can be used to reconstruct the previous images which is shown in figure 13.

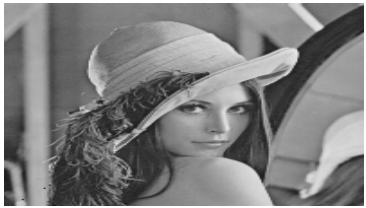

Fig 13: Reconstructed image

## V. Experimental Result

Figure 14 showing the experimental result after embedding a secret message using the technique described in section 3.

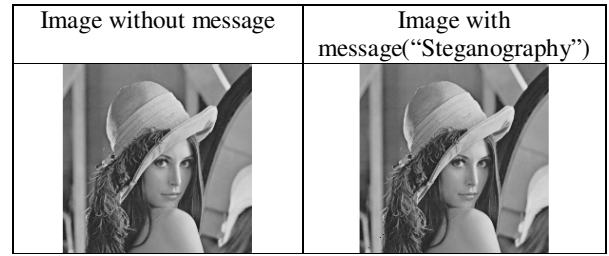

| Image without message | Image with message("Steganography") |
|---|---|

Fig 14: image before and after embedding secret message.

Dfference distortion metrics and correlation distortion metrics are shown in table 2 and table 3 shows histograms of stego image and reconstructed stego image.

TABLE 2

DFFERENCE DISTORTION METRICS AND CORRELATION DISTORTION

| Difference Distortion Metrics ||
|---|---|
| Metrics | Value |
| Maximum Difference | R=33,G=39,B=35 |
| Average Absolute Difference | R=.013025,G=0.0132,B=0.013375 |
| Norm Average Absolute Difference | R=2.625638E-09,G=2.660915E-09, B=2.696193E-09 |
| Mean Square Error | R=1.738725,G=1.82925,B=1.784725 |
| Normalized Mean Square Error | R=9.86075370883913E-05, G=0.00010376829743219, B=0.000101242517225456 |
| SNR | R=10141.2126245847,G=9636.85465354654 B=9877.27317934136 |
| PSNR | R=27.6888315189939,G=26.3117690904355 B=26.968190398341 |
| Image Fidelity | R=0.999901392462912,G=0.999896231702568 B=0.999898757482775 |
| Correlation Distortion Metrics ||
| Normalized Cross-correlation | R=0.999892545850374,G=0.999889763379422 B=0.99989083126965 |
| Correlation Quality | R=142.127729580045,G=142.127334071267 B=142.127485864391 |

TABLE 3

HISTOGRAM

| Image with embedded message before applying the method of visual cryptography | Reconstructed stego image |
|---|---|
| 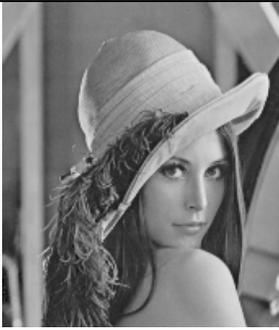 | 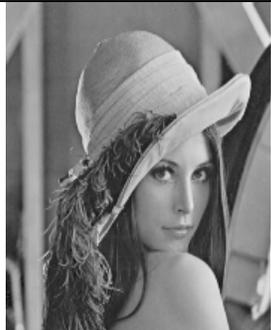 |
| Histogram | Histogram |
| 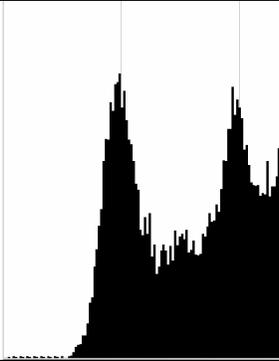 | 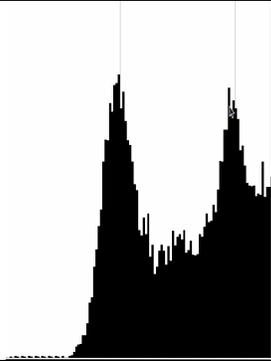 |

TABLE 4

SHOWING SNR AND PSNR FOR DIFFERENT IMAGES

| | 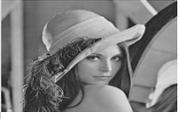 | 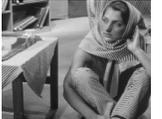 |
|---|---|---|
| | LENA | BARBARA |
| SNR | R=10141.2126245847 G=9636.85465354654 B=9877.27317934136 | R=58884.6180223285 G=52250.318342637,B= 45662.4844716229 |
| PSNR | R=27.6888315189939 G=26.3117690904355 B=26.968190398341 | R= 170.58587137579 G=151.366628221335 B= 132.281994252244 |
| | 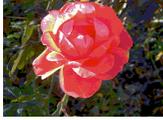 | 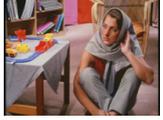 |
| | Rose | BARBARA (COLOR) |
| SNR | R=149125.445855116 G= 22049.939301848 B=22832.3875521031 | R=1466745.08525346, G=997857.042713568 B=3314268.88990826 |
| PSNR | R= 359.453350326207 G= 158.129950254387 B= 182.382734401967 | R=3548.8150126215 G=3869.81335547169 B=14130.1441786948 |

A. *Comparative Results for Different Images*

Table 4 is showing the SNR and PSNR values for different images .

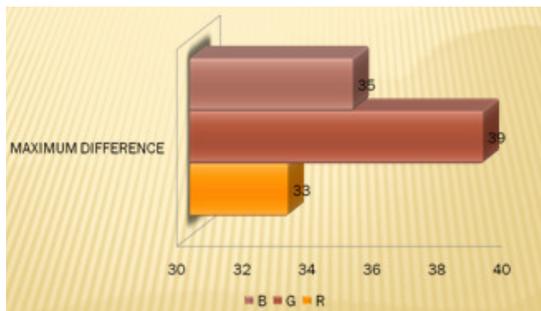

Fig 15(a) showing maximum difference (MD)

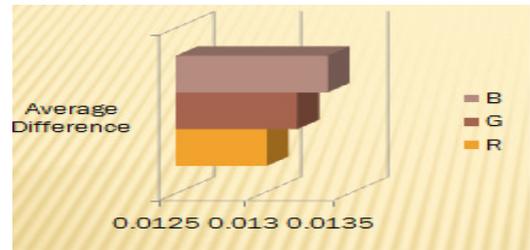

Fig 15(b) showing Average difference (AD)

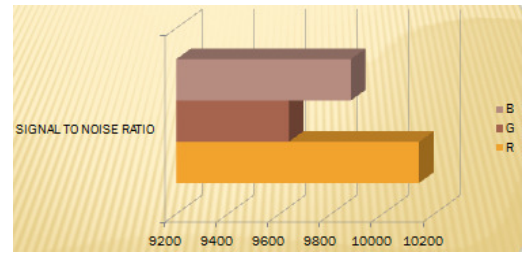

Fig 15(c) showing SNR

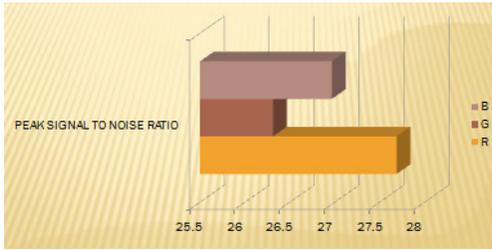

Fig 15(d) showing PSNR

Fig 15: Graphical representation distortion metrics

Figure 15 graphical representation of different distortion metrics.

*B. Comparison with other methods*

The paper developed by Chetana Hegde, Manu S, P Deepa Shenoy, Venugopal K R, L M Patnaik [11], is on visual cryptography. Let us compare the characteristic features of the proposed algorithm with the algorithm developer by Chetana Hegde et. al.

TABLE 5

COMPARISON RESULT

| Features | Algorithm proposed by Chetana Hegde et. al. [11] | Proposed Algorithm |
|---|---|---|
| Type of the algorithm | (2,2) visual cryptography | (4, n) visual cryptography |
| Pixel expansion | 4 times | variable |

Table 1 presents the comparison of some salient features between the proposed algorithm and the algorithm proposed by Chetana Hegde et. al. [4] .

## VI. CONCLUSION AND FUTURE WORK

In this paper a novel (4, n) visual cryptographic technique has been proposed which is easy to implement and it is capable to provide security like other existing algorithms. Above all, the proposed algorithm is a simple, straight-forward but intrinsically strong and compact approach to visual cryptography using the essence of steganography operations. A comparative study and security level will be verified in future with other well known algorithms.